\documentclass[prl,showpacs,twocolumn,floatfix]{revtex4} 
\usepackage{epsfig} 
\usepackage{mathrsfs} 
\usepackage{amssymb} 
\usepackage{color} 
\usepackage{graphics}
\usepackage{psfrag}
\usepackage{graphicx} 
\newcommand{\ecen}{\end{center}} \newcommand{\btab}{\begin{tabular}} 
\newcommand{\etab}{\end{tabular}} \newcommand{\bdes}{\begin{description}} 
\newcommand{\edes}{\end{description}}  
 \newcommand{\beq}{\begin{equation}} 
\newcommand{\eeq}{\end{equation}} \newcommand{\bea}{\begin{eqnarray}} 
\newcommand{\eea}{\end{eqnarray}}  
  
\newcommand{\bary}{\begin{array}} \newcommand{\eary}{\end{array}} 
\newcommand{\benum}{\begin{enumerate}} 
\newcommand{\eenum}{\end{enumerate}} \newcommand{\bitem}{\begin{itemize}} 
\newcommand{\eitem}{\end{itemize}} 

\begin{document}


\title{Physics of interface: Strongly Correlated Barrier with chemical
modulation sandwiched between two metallic planes.}


\author{Tribikram Gupta } \email[]{trbkrm01@yahoo.co.in}
\affiliation{Theoretical Physics Division,
Indian Association for the Cultivation of Sciences, Kolkata, India}
\author{Sanjay Gupta} \email[]{sanjay1@bose.res.in}
\affiliation{  Department of Chemical, Biological and Macromolecular Sciences,
S. N. Bose National Centre for Basic Sciences, Kolkata, India}


\date{\today}

\begin{abstract}

Barrier planes described by the Ionic Hubbard model and sandwiched between
metallic planes on both sides are studied using unrestricted Hartree Fock. For 
zero onsite correlation, the presence of the metallic interface generates an additional 
gap in the energy spectrum away from half filling, if the chemically modulated potential 
in the barrier planes exceed a critical value. There is reentrant behaviour and an 
insulator-metal-insulator transition as we tune onsite correlation for fixed site potential.
The metal is thus able to penetrate the barrier planes due to proximity effect, in a system 
that is otherwise an insulator throughout.


\end{abstract}


\maketitle

\section{Introduction} 
The emergence of unexpected new properties has made 
the physics of interface an active research area for some time 
now\cite{Ohtomo,Millis1,Millis2,Thiel,Kenji,Dagotto1,Rosch,Krish}.
The interface of a band insulator and Mott insulator was shown to 
have metallic properties\cite{Ohtomo}, which becomes superconducting
on lowering the temperature\cite{Thiel,Kenji}. Theoretically 
the subject has been studied using different methods like 
restricted Hartree Fock\cite{Millis1}, two site DMFT\cite{Millis2},
single site DMFT \cite{Rosch, Krish} and Lancsoz \cite{Dagotto1}. 

In this letter we have studied a metal-quasi 2D barrier-metal
heterostructure. The barrier planes are described by the Ionic 
Hubbard Model, which describes a band insulator with onsite correlation.
Thus the barrier planes describe a band insulator in the absence of correlation $U$,
and a Mott insulator in the absence of chemical modulation for $U \ge U_c$,
where $U_c$ depends on the dimensionality of the system.  
The ionic Hubbard model has been widely studied by 
various groups using various techniques like single site DMFT\cite{Arti,Craco},
quantum monte carlo\cite{Scalettar} and cluster DMFT \cite{Dagotto2} in 2D, 
DMRG in 1D\cite{Baeriswyl} and two site DMFT in quasi 2D \cite{Byczuk}.
At half filling a single site DMFT solution which assumes a paramagnetic solution yields 
a metallic phase which separates the Mott insulator and band insulator phases\cite{Arti,Craco}.
A solution which incorporates long range anti-ferromagnetic ordering in quasi 
2D, using two site DMFT, however shows that the system is insulating for 
all interaction strengths.

The Mott insulator-band insulator heterostructure has been studied using 
restricted Hartree Fock \cite{Millis1}, inhomogeneous two site DMFT \cite{Millis2}
and in quasi one dimensional lattices using Lancsoz method \cite{Dagotto1}. The 
metal-Mott insulator-metal heterostructure has been studied \cite{Rosch,Krish,Gupta2}
while the role of disorder on the Mott planes has also been investigated by us
\cite{Gupta3}. 

Spatial variations in the planes and along the z direction play a crucial 
role in determining the correct ground state of a highly inhomogeneous 
systems like heterostructures. The in plane variation is totally ignored 
in a one site inhomogeneous DMFT(IDMFT)\cite{Georges,Metzner,Potthoff,freericks_idmft} 
approach. The two site IDMFT 
method though capable of capturing planar long range order like 
planar antiferromagnetism, is insufficient to capture the regime where 
the two orders start to cancel each other leading to short range ordering. With 
this in mind we  have adopted the method of unrestricted Hartree Fock. The 
Hartree Fock method has been used successfully to describe the ground state of 
the ionic Hubbard model even in 1D and was shown to be in excellent agreement 
with real space RG calculation\cite{Sanjay}.   

\subsection{Model and Method} 
The Hamiltonian for the system is
\begin{eqnarray} 
\mathcal{H}&=&-\sum_{ij\alpha\sigma}t_{ij}^\parallel 
c^\dagger_{i\alpha\sigma}c^{}_{j\alpha\sigma} - t \sum_{i\alpha\sigma} 
[c^\dagger_{i\alpha\sigma}c^{}_{i\alpha+1\sigma}+h.c.]\nonumber\\ 
&-&\mu\sum_{i\alpha\sigma}c^\dagger_{i\alpha\sigma}c^{}_{i\alpha\sigma} 
+\sum_{i{\in B},\alpha}W_\alpha(n_{i\alpha\uparrow} + n_{i\alpha\downarrow})\nonumber\\ 
&+&\sum_{i\alpha}U_\alpha(n_{i\alpha\uparrow} 
-\frac{1}{2})(n_{i\alpha\downarrow}-\frac{1}{2}).\; 
\end{eqnarray} 
Here the label $\alpha$ indexes the planes, and the label $i$ 
indexes sites of the two-dimensional square lattice in each plane. 
The operator $c^\dagger_{i\alpha\sigma}$ ($c^{}_{i\alpha\sigma}$) 
creates (destroys) an electron of spin $\sigma$ at site $i$ on the 
plane $\alpha$. We set the in-plane hopping $t^{\parallel}$ to be 
nearest neighbor only, and equal to $t$, the hopping between planes, 
so that the lattice structure is that of a simple cubic lattice.  
We take $W_\alpha = W$ for the $B$ sites and $W_\alpha = 0$ for 
the $A$ sites. We also take 
$U_\alpha = U$ for the barrier planes, and zero for the metallic 
planes. The chemical potential $\mu$ is calculated by taking 
the average of the N/2 th and the N/2 + 1 th energy level. 
The plane index $\alpha = 1, m$ correspond to the metallic layers. All the other 
$\alpha$ values in between correspond to the barrier planes.
For the quasi 2D geometry taken by us total number of sites 
$N = m\times{L^2}$, where $m$ is the total number of layers 
and $L^2$ is the number of sites in the planes.

\begin{figure}
\begin{center}
   \begin{tabular}{cc}
      \resizebox{39.5mm}{!}{\includegraphics{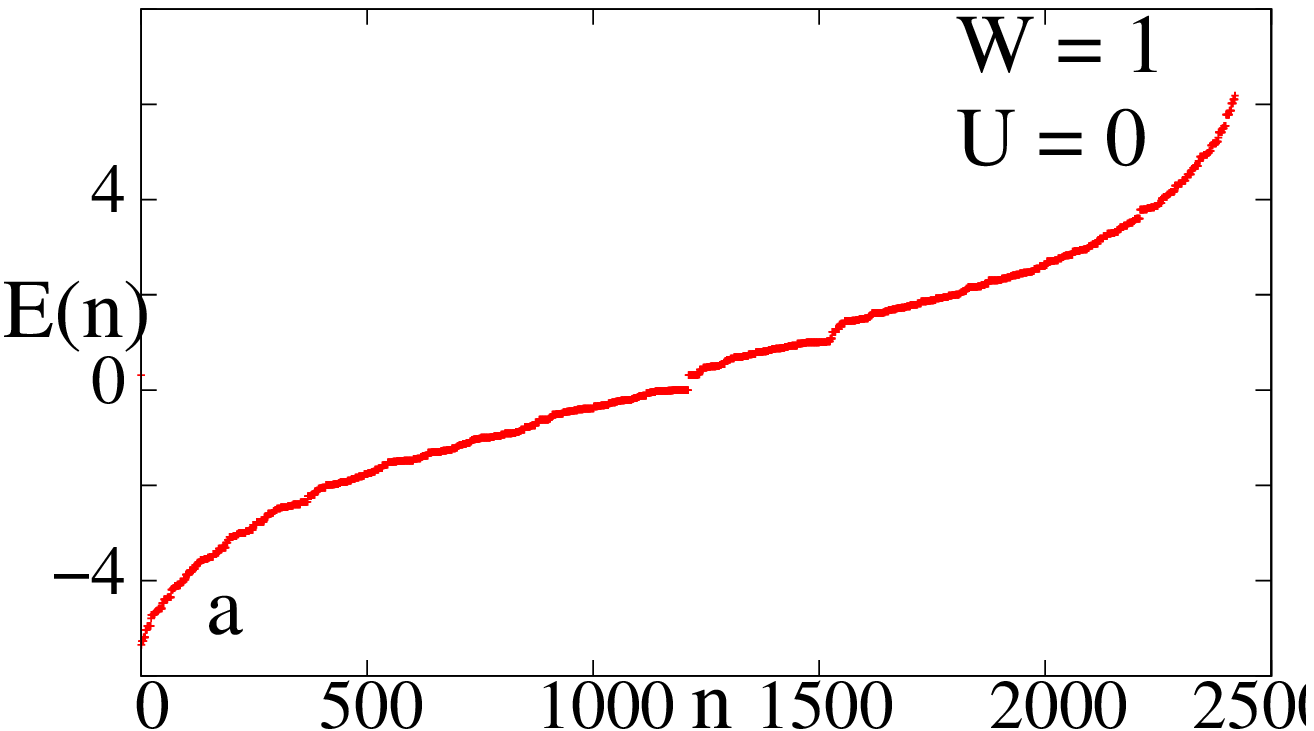}} &
      \resizebox{39.5mm}{!}{\includegraphics{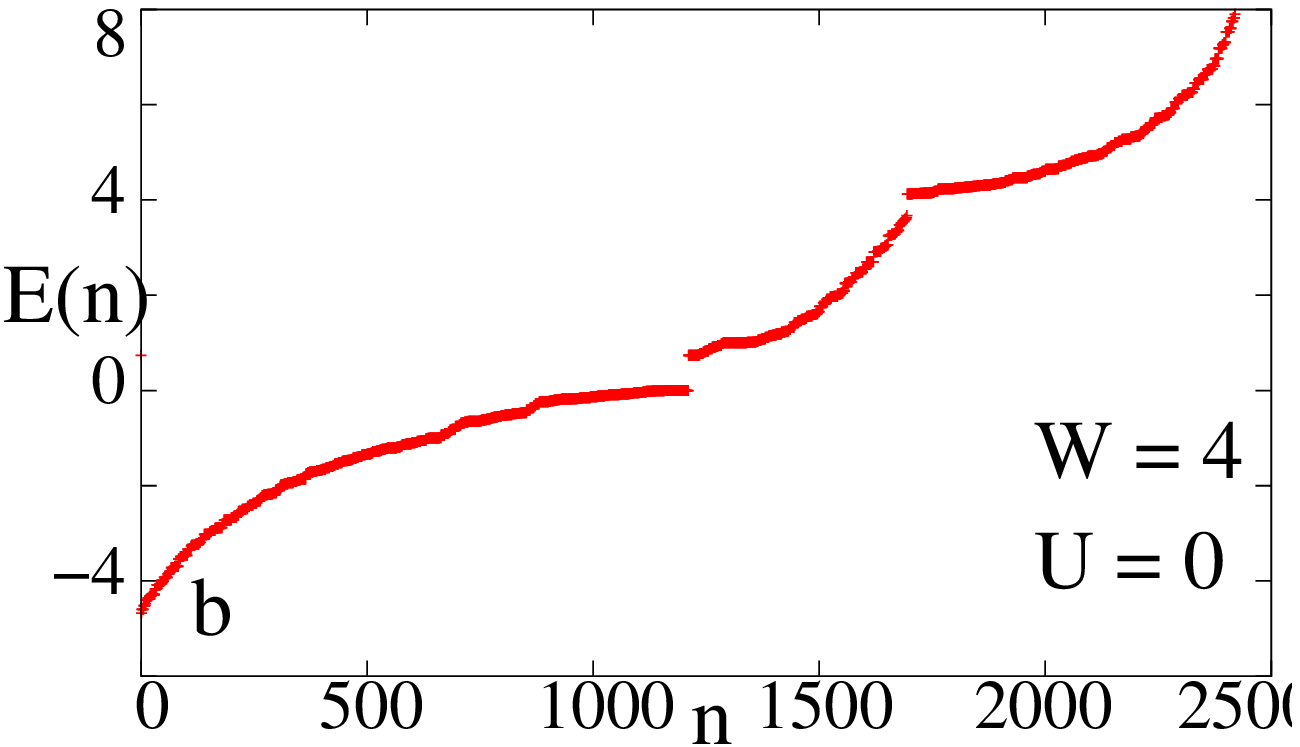}} \\ 
      \resizebox{39.5mm}{!}{\includegraphics{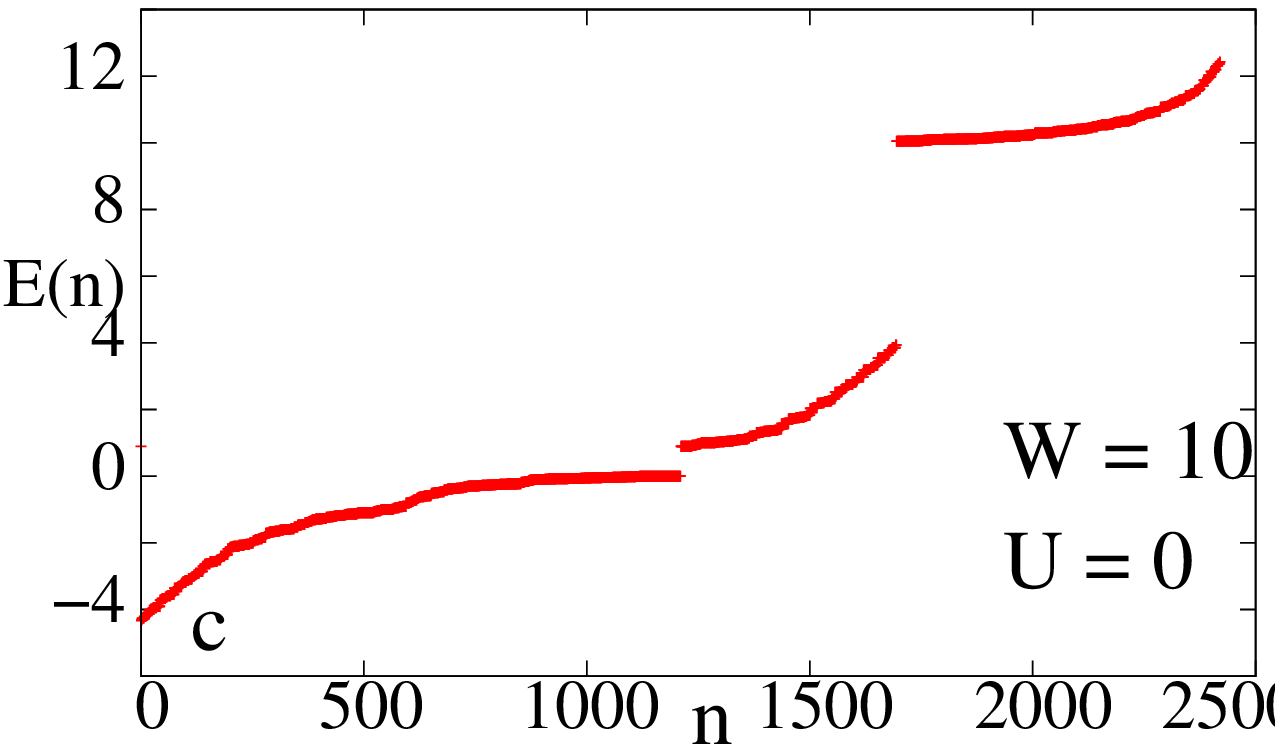}} &
      \resizebox{39.5mm}{!}{\includegraphics{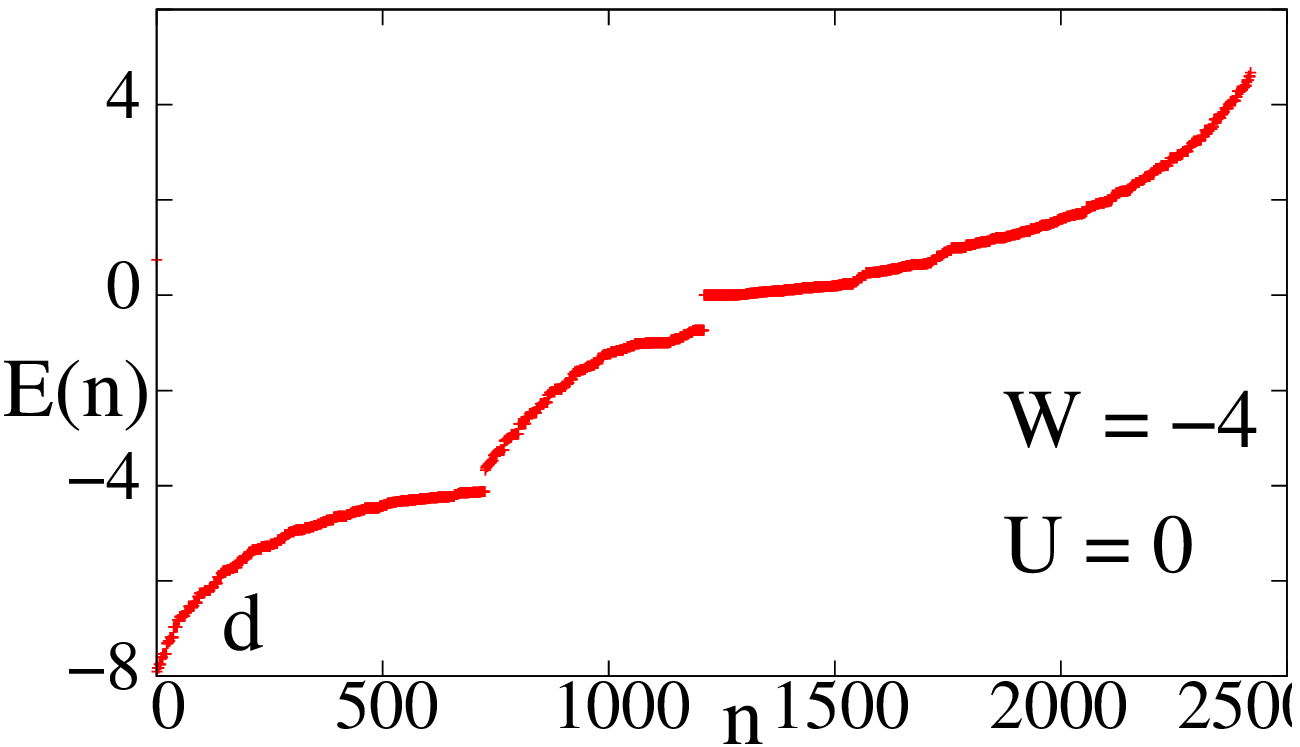}} \\ 
      \resizebox{39.5mm}{!}{\includegraphics{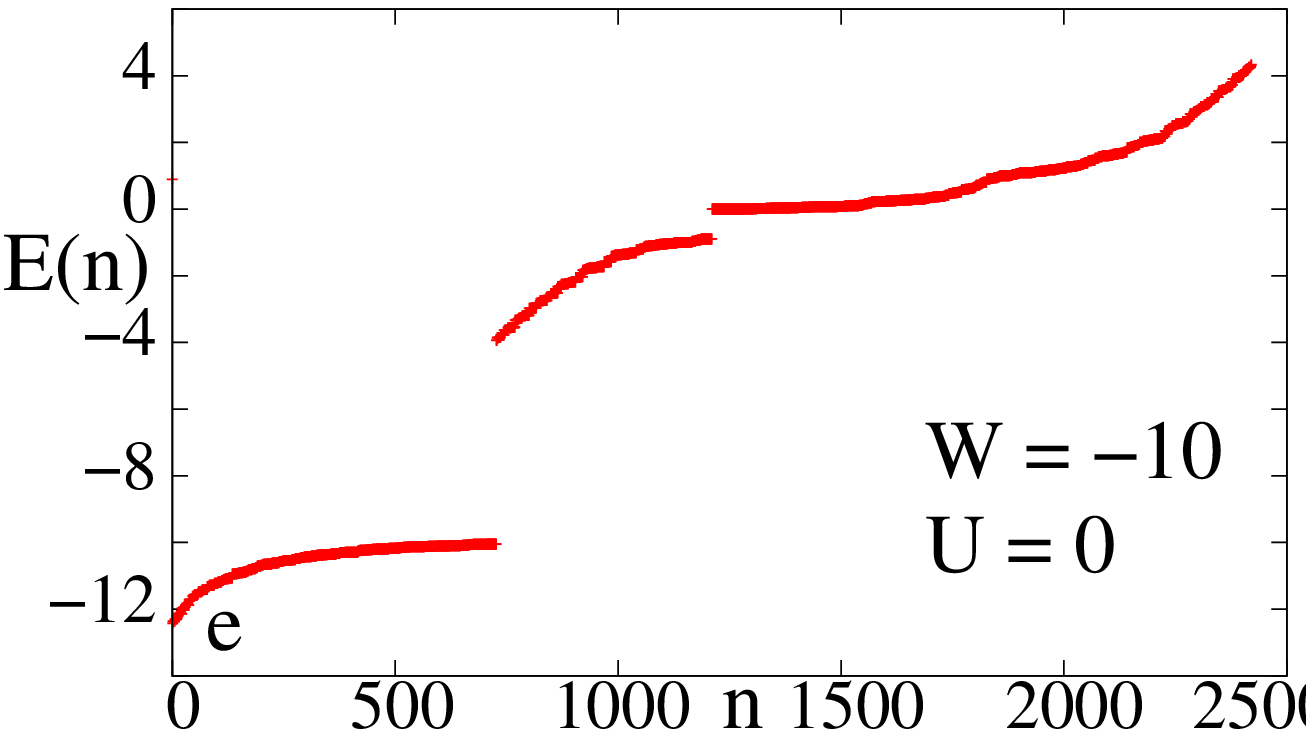}} &
      \resizebox{39.5mm}{!}{\includegraphics{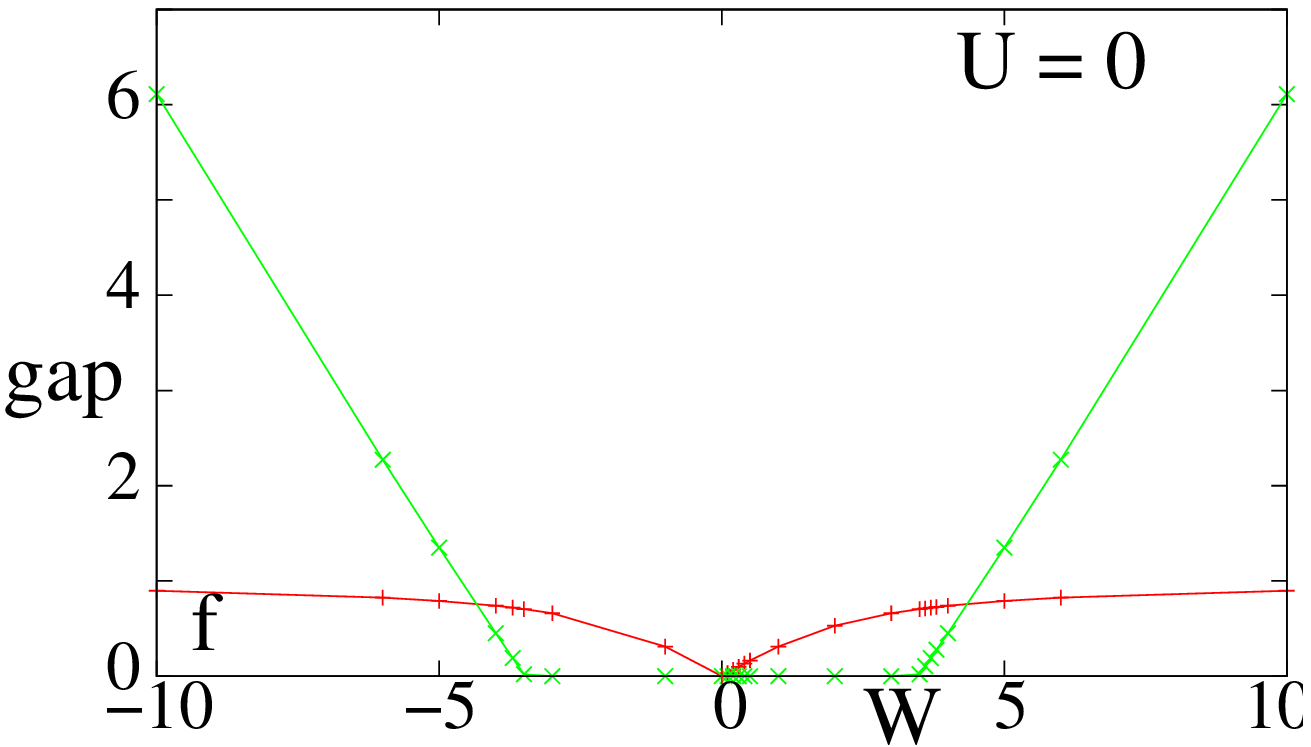}} \\ 
    \end{tabular}
\caption{Energy spectrum for $U = 0$ for a) $W$ = 1, b) $W$ = 4, c) $W$ = 10,
d) $W$ = -4, e) $W$ = -10. f) Evolution of the multigaps with increasing $W$ 
for $U = 0$.}
\label{fig;GapatHalfFilling}
\end{center}
\end{figure}

\subsection{Calculated quantities} 
We have calculated the energy spectrum,  
charge and spin profile, and the optical conductivity. 
The charge at site $i$ is given by:
$C_i$ = $n_{i,\uparrow}$ + $n_{i,\downarrow}$. The spin at a 
particular site is given by $S_i$ = $|(n_{i,\uparrow} - n_{i,\downarrow})|$.
The optical conductivity is calculated using the Kubo formula. In this paper, 
$\sigma_{zz}(\omega)$ is  calculated in units of ${\pi  e^2 }/{{\hbar a_0}}$.
For details of the calculation procedure see our previous work\cite{SanjeevEPL,Gupta1}.
We have performed finite size scaling of each of our transport results. 
The value of $\sigma_{zz}(\omega)$ is calculated for multiples of a small $\omega = \omega_r$, 
where $\omega_r$ is twice the lowest $\omega$ that we can access for a particular system size.

\begin{figure}
\begin{center}
   \begin{tabular}{cc}
      \resizebox{39.5mm}{!}{\includegraphics{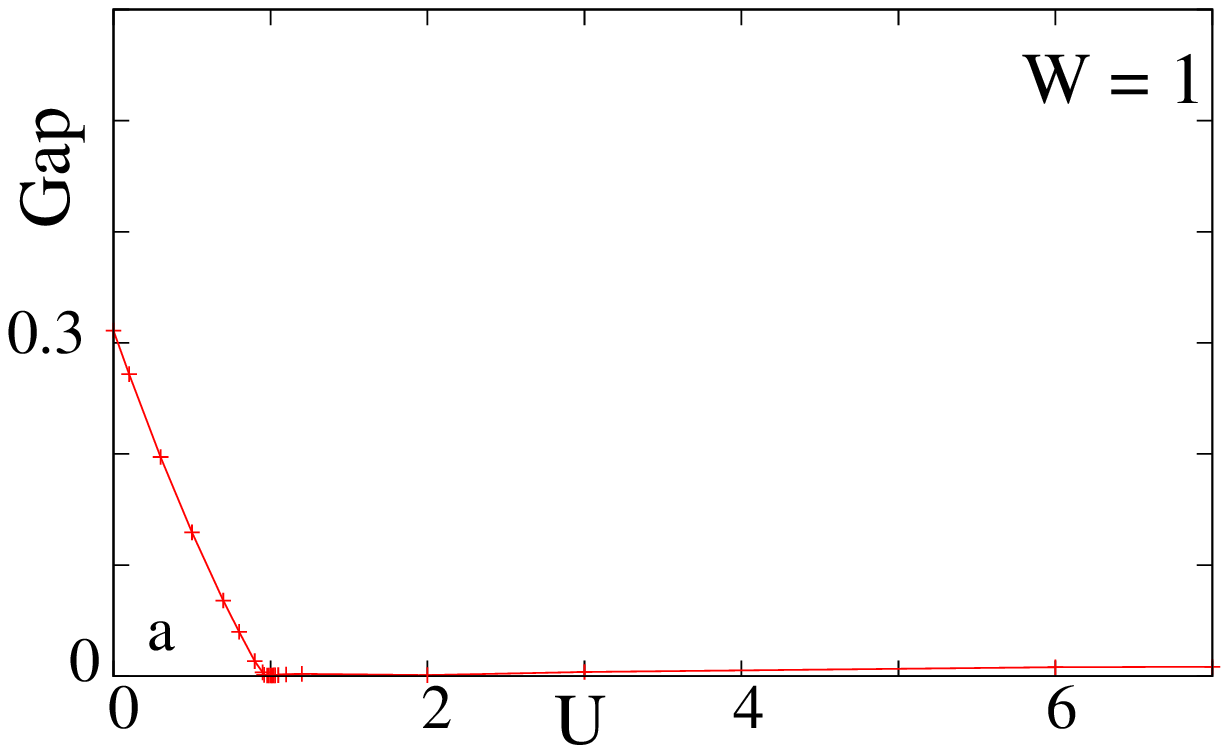}} &
      \resizebox{39.5mm}{!}{\includegraphics{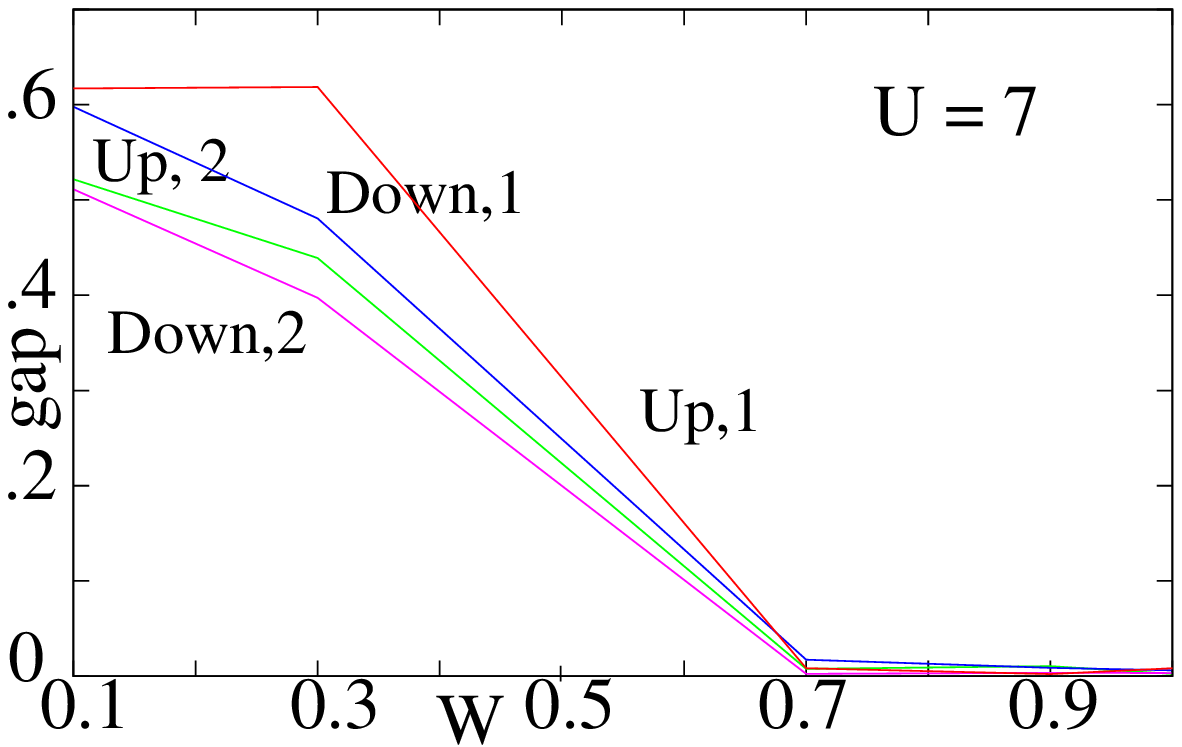}} \\ 
      \resizebox{39.5mm}{!}{\includegraphics{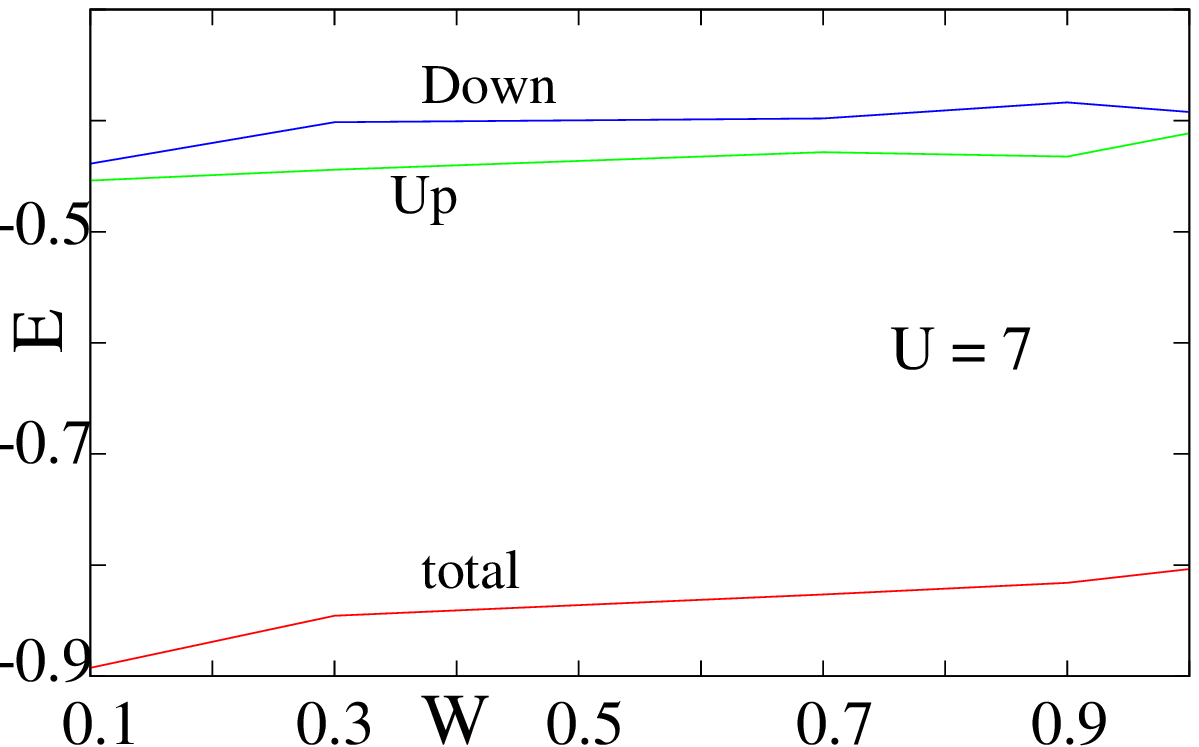}} & 
      \resizebox{39.5mm}{!}{\includegraphics{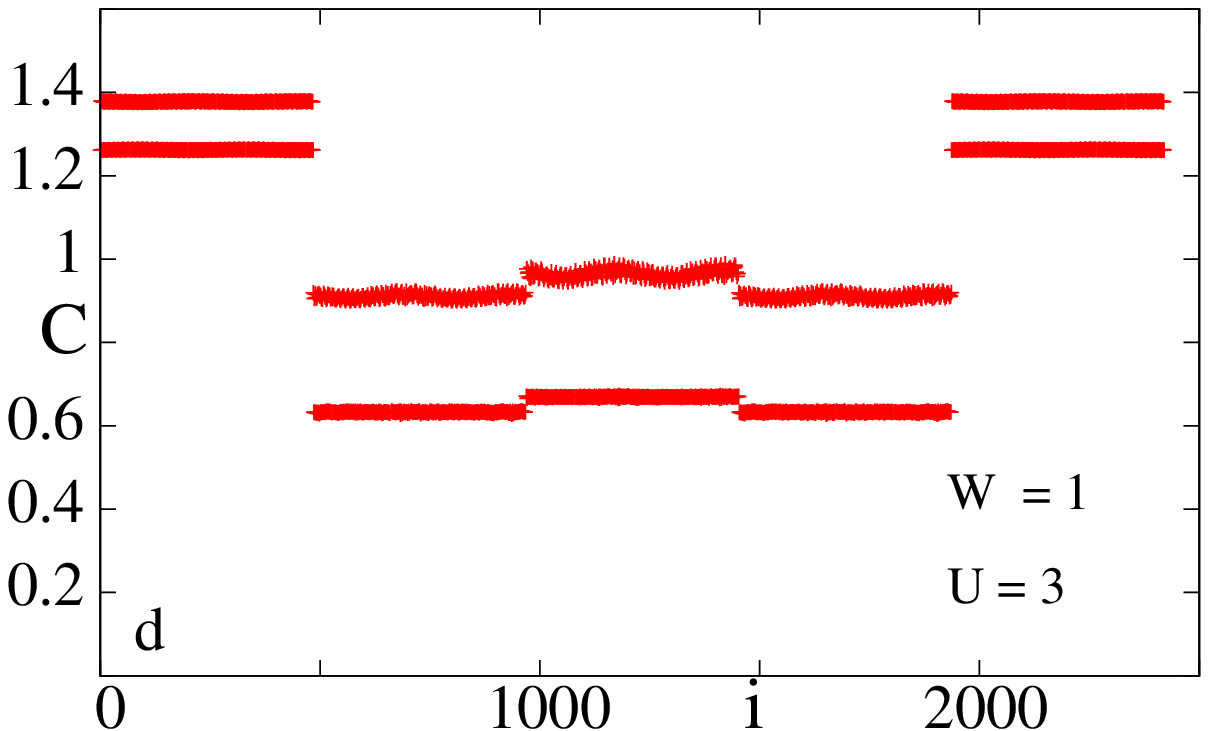}} \\ 
    \end{tabular}
\caption{a. Gap at half filling for $W = 1$ with increasing $U$. There is a 
prolonged gapless phase that extends from $U = 1$ to $U = 7$. 
b. Gap at and away from half filling for both spins at $U = 7$ with $W$.
c. Energy of the two spins plotted separately along with the total energy.
d. Charge profile at $W = 1$ and $U = 3$. 
} 
\label{fig;Gap&Profile_W1_U3}
\end{center}
\end{figure}

\subsection{Analysis of our results} 

Fig 1a,b,c,d,e and f shows that different types of sites emerge  
as we systematically tune $W$. $W$ positive/negative corresponds 
to repulsive/attractive potential/wells in the barrier planes. 
In the absence of $U$ any non-zero $W$ leads to a 
charge density wave(CDW) profile, with a symmetry in the gap about 
the origin. Increasing $W$ beyond a certain threshold value, an 
additional gap opens up away from half filling. 
The value of the gap which would have been identical to $W$ in the 
absence of the metallic planes gets reduced in the presence of the metallic planes. This 
gap is due to the fact that the barrier planes induce a density modulation 
even in the metallic planes due to proximity effect. Thus the entire system 
has a rather inhomogeneous alternate density modulation.  
As we increase $W$ further at around $W_g = 4$, a second gap opens up. 
For general $m$, the filling at which the second gap opens up is given 
by  0.5 + sign(W)*(1/m). Above this value $W_g$, the lowly occupied 
sites in the barrier planes(B sites) become very different from the 
lowly occupied sites in the metallic planes and a new band emerges. 
For negative $W$ the second gap opens at $W$ = -4.
The states below this new band gap now
correspond to the B sites from the barrier planes and the 
states above the gap correspond to lowly occupied sites from the metallic
planes. Due to -ve value of $W$ the system can now lower energy by doubly 
occupying the $B$ sites. The gap at half filling saturates to a fixed
value as we keep on increasing $W$. This is because the higher energy 
B sites(both in the barrier and those induced in the metallic planes) 
are now more or less avoided by the electrons.    

In Fig 2a, we show the modulation of the gap at half filling 
for $W$ = 1 as we increase $U$. The band gap is suppressed by 
$U$ down to zero, after which there is a large gapless region.
The low $U$ region corresponds to a CDW phase, where $U$ has started to compete with the 
chemical modulation with the latter being dominant. The large gapless region corresponds to 
a highly inhomogeneous phase, with no spin order till $U = 6$ and with gradually 
diminishing but non zero charge order. Above $U = 6$, spin order starts to pick up 
gradually, though there is still a remnant weak charge order also.  

\begin{figure}
\begin{center}
   \begin{tabular}{cc}
      \resizebox{39.5mm}{!}{\includegraphics{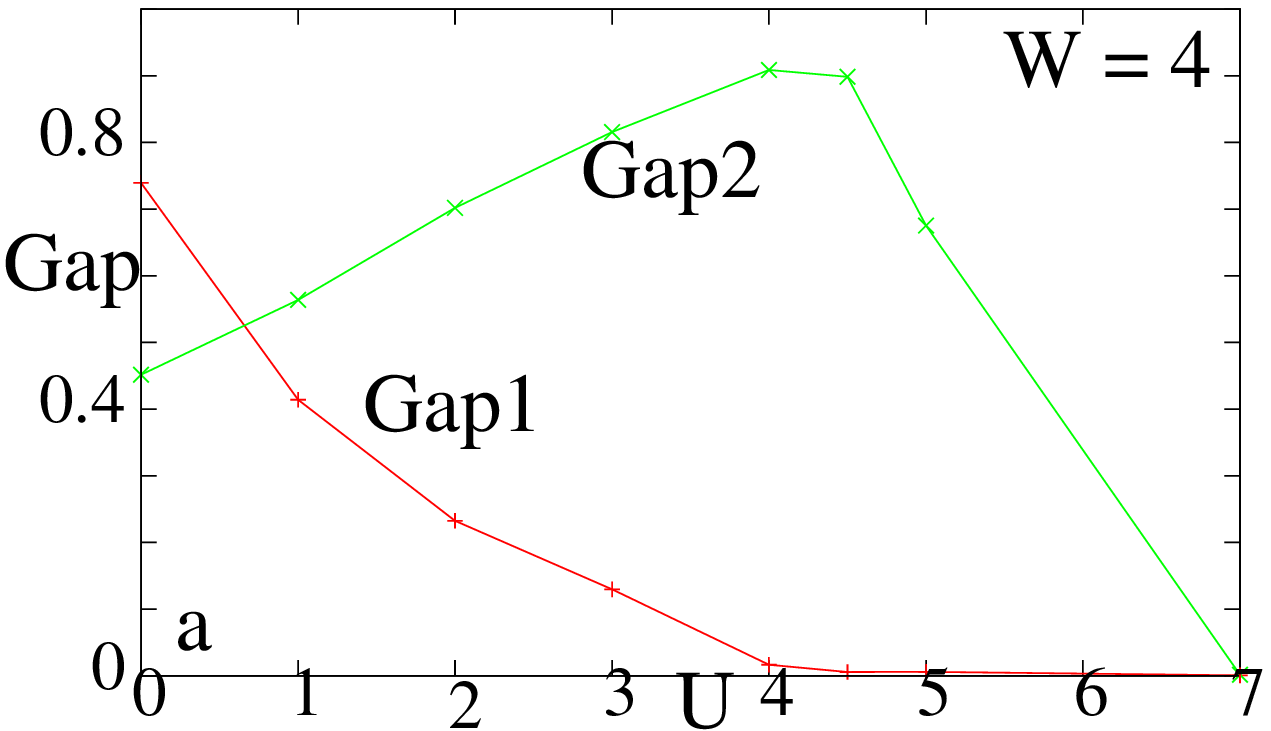}} &
      \resizebox{39.5mm}{!}{\includegraphics{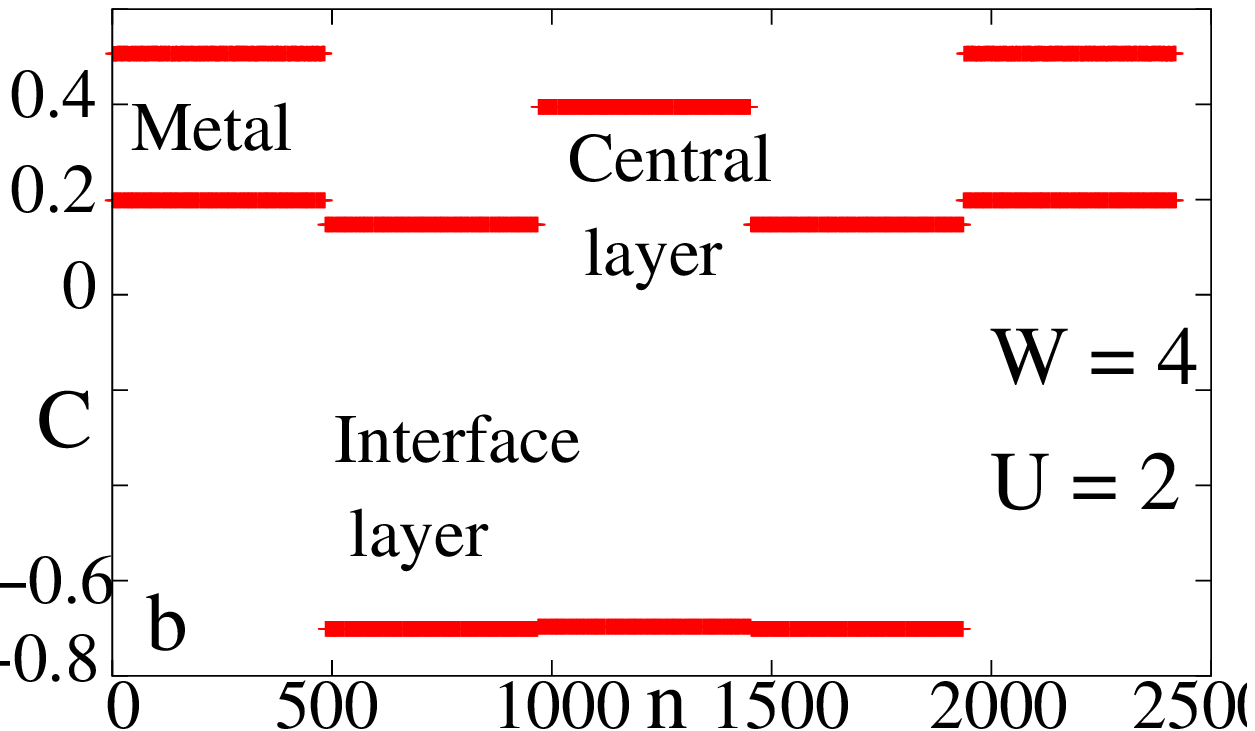}} \\ 
      \resizebox{39.5mm}{!}{\includegraphics{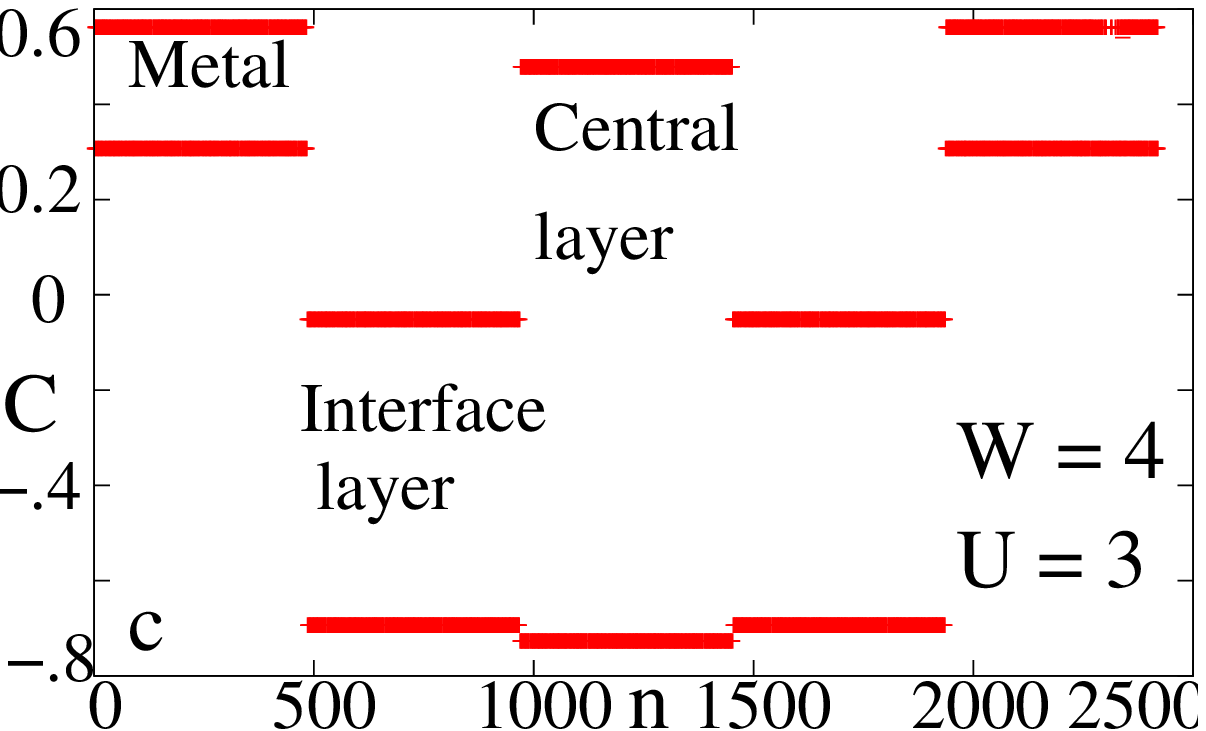}} & 
      \resizebox{39.5mm}{!}{\includegraphics{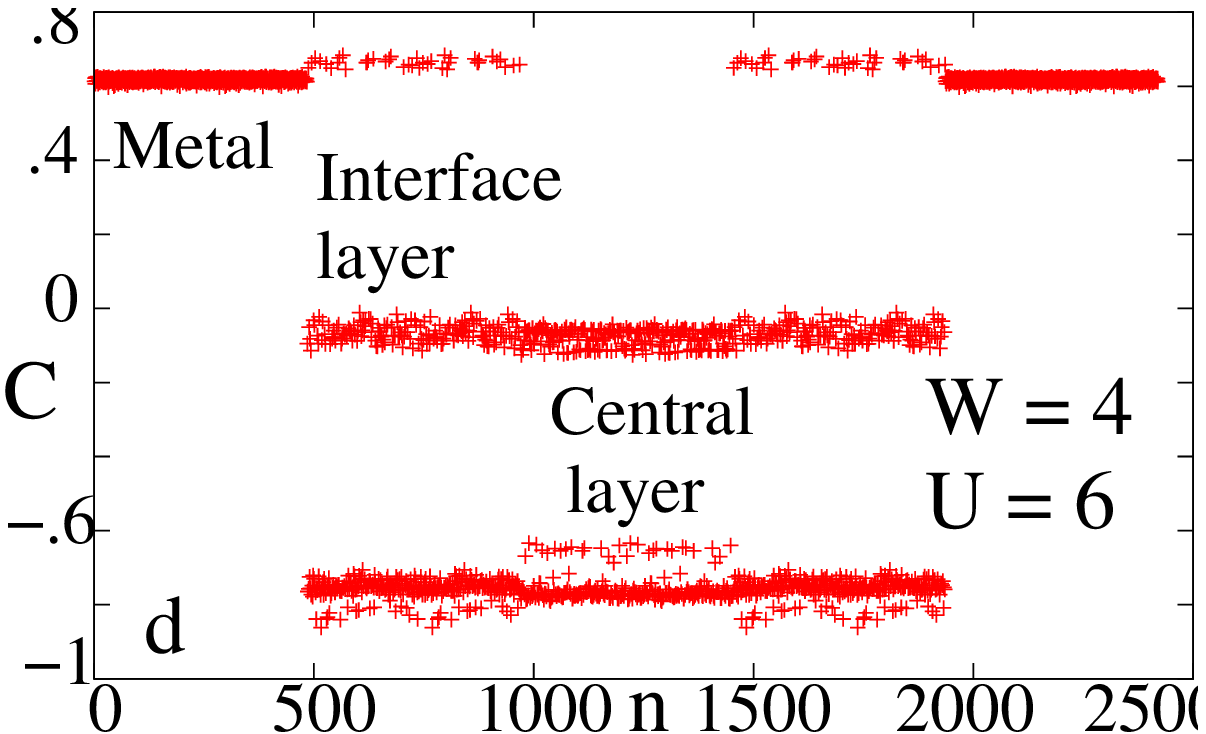}} \\ 
    \end{tabular}
\caption{a. Gap at and away from half filling for $W = 4$ with increasing $U$. 
b,c,d. Charge profiles for $W = 4$ and $U = 2,3,6$ respectively.
}
\label{fig;Gap&IPR&Profile_W_4}
\end{center}
\end{figure}

Fig 2b shows the variation of the two gaps in the energy spectrum 
for $U$ = 7, for both spins with variation in $W$. In the absence of $W$, the opening up 
of two gaps in the energy spectrum was first shown by us in a previous work
\cite{Gupta2}. For positive values of $W$, till $W = 1$, both the gaps for both spins 
fall monotonically with increasing $W$. There is small spin asymmetry which is a signature 
of spontaneous symmetry breaking of the up and down spin sectors for such a system. This 
behaviour has already been observed and reported by Cabib and Callen \cite{C&C} for an 
extended Hubbard ($t-U-V$) model. 

While for positive $W$ the $B$ sites are avoided by the electrons,
The repulsion of the electrons to the metallic planes takes place 
uniformly for both up and down spins because there is very small charge modulation 
in the metallic planes and since there is $U$ on all the barrier plane sites, they would 
all like to push electrons away to the metallic planes, without paying the cost of $U$. 

  
Fig 2c shows the plots of energy for both spin species and the total energy 
as we modulate $W$ for $U = 7$. The total energy increases monotonically with 
increasing $W$. The energy for the two spin species are different for the 
entire parameter range where there is both charge and spin order. This is a 
signature of the spin asymmetry reported earlier\cite{C&C}. 

2d shows the plots of the charge profile for $U = 7$. The overall symmetry 
of the profiles about the central plane and the explicit z dependence are seen. 
In this regime where $W$ has just killed the gap due to anti ferromagnetic 
long range order due to $U$, there is already a strong plane dependent 
CDW ordering.


In Fig 3a we have shown the variation of the two gaps that opens up 
for $W$ = 4 with increase in $U$. We see that the two gaps behave in 
very contrasting manner with increasing $U$. While the gap at half 
filling falls off monotonically to zero, which basically implies 
that the charge ordering induced in the metallic
 planes by $W$ is killed by $U$. 
However the second gap for positive $W \ge 4$ shows 
a non monotonic variation with $U$. It first increases with 
increasing $U$ and then after a certain value of $U$ starts 
to decrease. This is because for low values of $U$, the electrons 
in the A sites of the interface layer start to deplete by moving to the metallic planes. 
The occupancy of the B sites however is relatively unaffected in this regime. Thus 
the second gap which senses the difference between the topmost level of the 
second band and the lowest level in the topmost band increases. 
However beyond a certain threshold, the 
B sites in the barrier planes start to get occupied and then 
the gap starts falling. 

Figs 3b,c,d show the charge profile for $U$ = 2,3,6 respectively for $W$ = 4.
While for $U$ = 2 and 3, the gap at half filling decreases but is still 
non zero, for $U$ = 6 the gap at half filling has already become zero. This 
we can see clearly from the figures, where for $U$ = 2 and 3, there is 
inhomogeneous CDW ordering in all the planes. For $U$ = 6 however the CDW 
ordering is destroyed in the metallic planes. 

\begin{figure}
\begin{center}
   \begin{tabular}{cc}
      \resizebox{39.5mm}{!}{\includegraphics{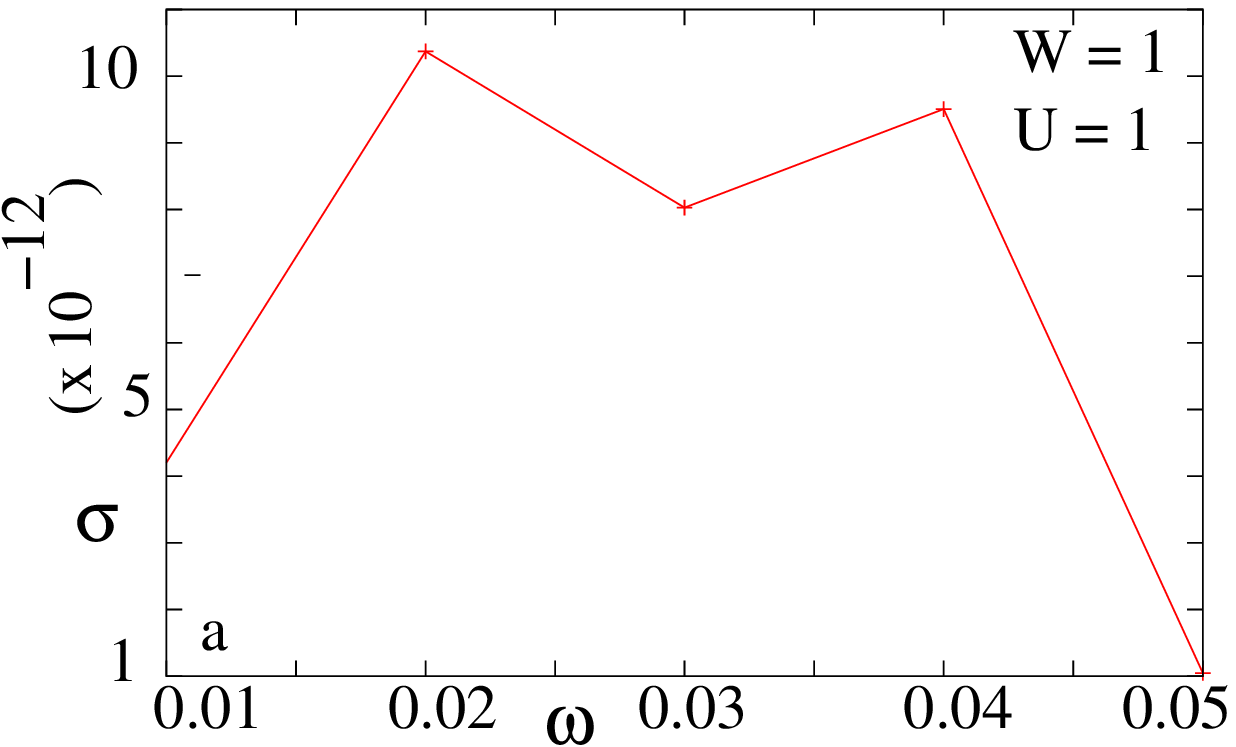}} &
      \resizebox{39.5mm}{!}{\includegraphics{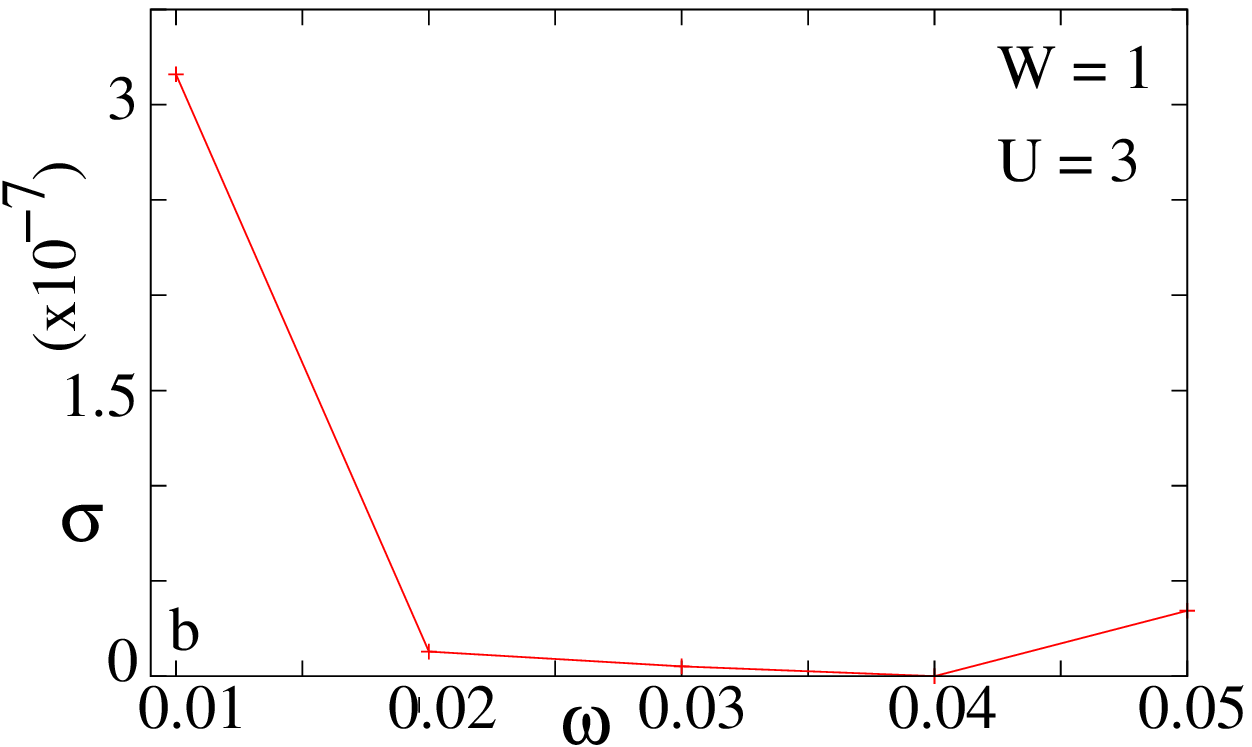}} \\ 
      \resizebox{39.5mm}{!}{\includegraphics{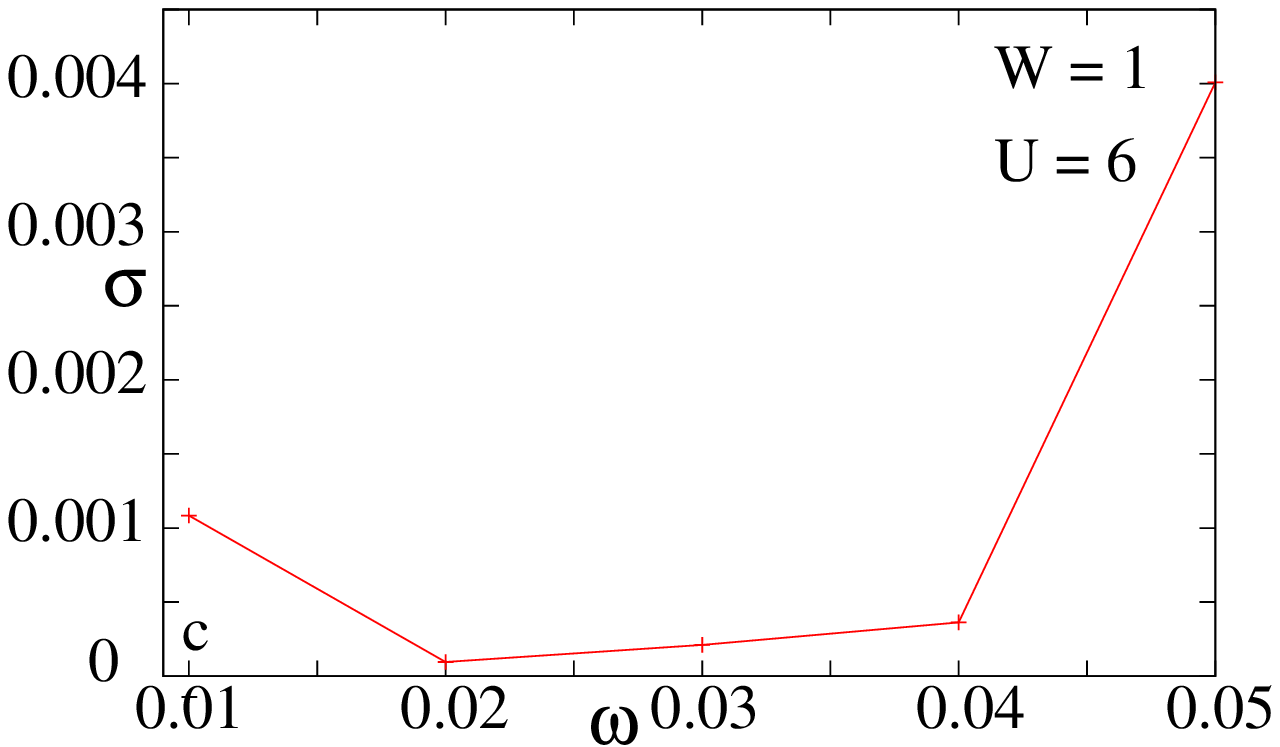}} &
      \resizebox{39.5mm}{!}{\includegraphics{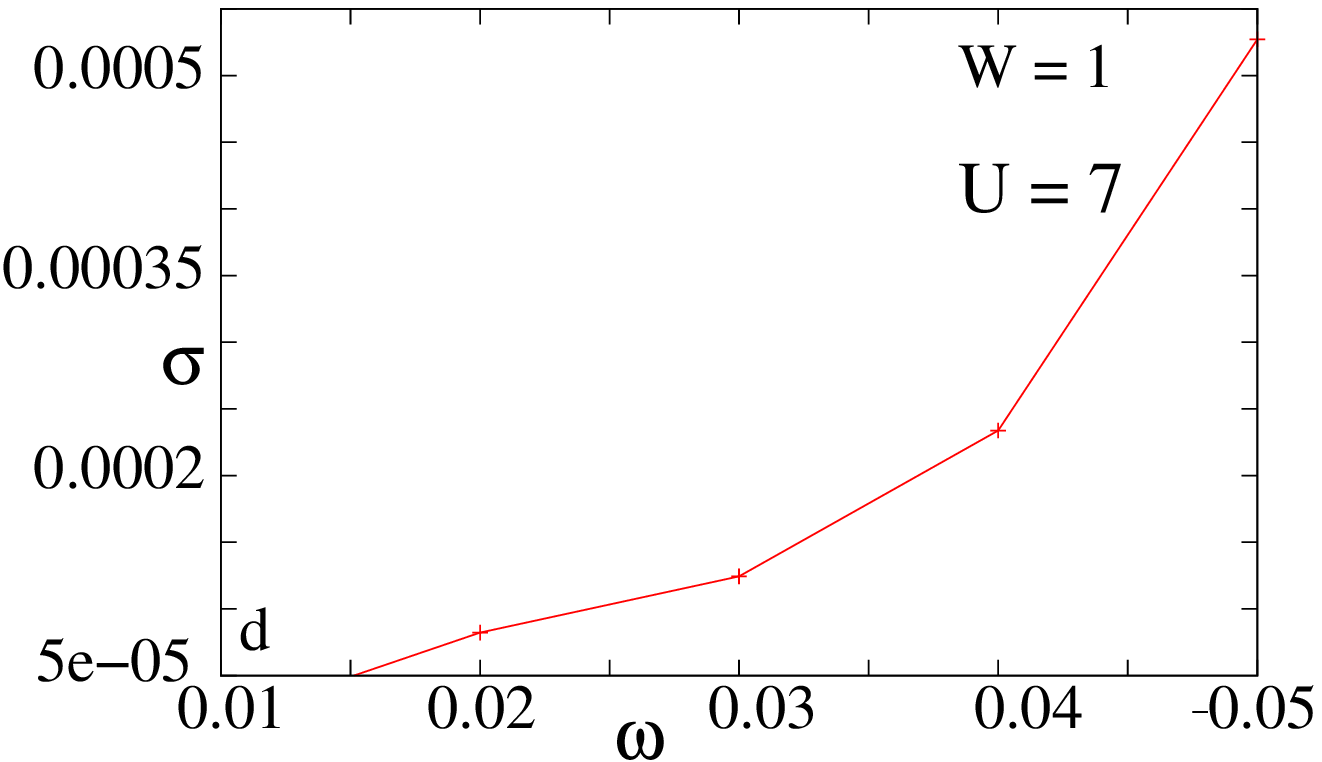}} \\ 
    \end{tabular}
\caption{Optical conductivity results $\sigma(\omega)$ vs $\omega$ 
for a. $U$ = 1, b. $U$ = 3, c. $U$ = 6, d. $U$ = 7.
}
\label{fig;Optical_Conductivity}
\end{center}
\end{figure}

Figs 4a,b,c,d show the plots of optical conductivity versus $\omega$ 
for $U$ = 1,3,6 and 7 respectively for $W$ = 1. 
As we tune $U$ keeping $W$ fixed, the system exhibits reentrant 
behaviour in the form of an insulator-metal-insulator transition. As the gap
at half filling gets killed initially around $U$ = 1, a residual slightly 
weakened CDW ordering persists in the system all the way upto $U$ = 6. 
The A sites in the barrier start to deplete and the B sites are relatively 
unaffected for low $U$.  
As we increase $U$ further, the system becomes a metal as seen in $U$ = 3 and 6.
The metal at $U$ = 3 is a much weaker one compared to the one at $U$ = 6, as the charge 
ordering is fully destroyed for $U$ = 6. The weak metallic behaviour 
for $U$ = 3 inspite of the presence of charge ordering in the system 
is a consequence of weak tunneling of electrons between the sites.
For $U$ = 7, a small inhomogeneous spin order develops in the system, which drives the system to 
an insulating phase.

\section{Conclusion} 
A barrier described by the ionic Hubbard model sandwiched between two 
metallic planes has been studied using unrestricted Hartree Fock. 
Increasing the strength of the site potential modulation, beyond a 
critical strength $W_c$ opens a new gap away from half filling at 
a value that depends on the sign of the modulation.
Below $W_c$ onset of correlation reduces the gap to zero gradually.
Above $W_c$ while the gap at half filling follows the same pattern, 
the new gap shows non monotonic behaviour with increasing $U$.
The system shows reentrant insulator-metal-insulator transition with 
increasing $U$ for fixed $W$ = 1. 
Spin asymmetric behaviour is noticed when both charge and spin order 
is present in the system.

\end{document}